\documentclass[reprint,amsmath,amssymb,aps,superscriptaddress,hidelinks]{revtex4-2}
\usepackage[utf8]{inputenc}
\usepackage{graphicx}
\usepackage{hyperref}
\usepackage{dcolumn}
\usepackage{bm}
\usepackage{tcolorbox}

\begin{document}

\title{Transient amplification in stable Floquet media}

\author{Ioannis Kiorpelidis}
\affiliation{LAUM, UMR-CNRS 6613, Le Mans Universit\'e, Av. O. Messiaen, 72085, Le Mans, France}
\affiliation{Department of Physics, University of Athens, 15784, Athens, Greece}

\author{Fotios K. Diakonos}
\affiliation{Department of Physics, University of Athens, 15784, Athens, Greece}

\author{Georgios Theocharis}
\affiliation{LAUM, UMR-CNRS 6613, Le Mans Universit\'e, Av. O. Messiaen, 72085, Le Mans, France}

\author{Vincent Pagneux}
\affiliation{LAUM, UMR-CNRS 6613, Le Mans Universit\'e, Av. O. Messiaen, 72085, Le Mans, France}

\begin{abstract}
The Mathieu equation occurs naturally in the description of vibrations or in the propagation of waves in media with a time-periodic refractive index. It is known to lead to exponential parametric instability in some regions of the parameter space. However, even in the stable region the matrix that propagates the initial conditions forward in time is non-normal, and therefore, it can result in transient amplification. By optimizing over initial conditions as well as initial time we show that significant transient amplifications can be obtained, going beyond the one simply stemming from adiabatic invariance. Moreover, we explore the monodromy matrix in more depth by studying its  $\epsilon$-pseudospectra and Petermann factors, demonstrating that is the degree of non-normality of this matrix that determines the global amplifying features. In the context of wave propagation in time-varying media, this transient behavior allows us to display arbitrary amplification of the wave amplitude that is not due to exponential parametric instability.
\end{abstract}

\maketitle


\section{Introduction}
\label{Section1}

Over the last few years, the modulation of the properties of materials in time has attracted great interest \cite{Caloz2019,Galiffi2022,Segev2023}. Time-varying metamaterials exhibit rich phenomenology, ranging from time reflection and time refraction \cite{Agrawal2014} to nontrivial topological features \cite{Segev2018}. When these modulations are periodic in time, the prism of Floquet analysis can be used, leading to the development of Floquet metamaterials \cite{Alu2022,Fan2022}. Meanwhile, Floquet theory captures the stability properties of the solutions in terms of the Floquet exponents, and it is known that unstable solutions are related to parametric resonances \cite{Fossen2012}. They appear in a wide range of time-varying systems (we note that parametric resonances may appear in nonoscillating systems as well \cite{Clerc2012}), for instance, in photonic time crystals \cite{Segev2022} and in elastic metamaterials \cite{Norris2018,Ruzzene2019}.

Amplification in time-varying media is closely related to the concept of parametric instability, but there are other ways to amplify a system. For example, a new mechanism for gain was recently found in time-dependent photonic metamaterials \cite{Pendry2021}, resulting from the compression of the lines of the electric and magnetic fields. Furthermore, it is known, especially in hydrodynamics \cite{Trefethen1993,Schmid2007}, that stable solutions of a system can be transiently amplified when the matrix that propagates the initial conditions forward in time is non-normal, thus having nonorthogonal eigenvectors \cite{Ioannou1996a,Ioannou1996b}. Along this line, the pseudospectrum tool was developed in order to describe these transient amplifying phenomena \cite{Trefethen2005}. Let us remark that non-normality and the pseudospectrum appear to play an important role in the emerging field of non-Hermitian topology, for both the time transient \cite{Midya2022, Sato2023} and non-Hermitian skin effect spectrum \cite{Sato2023, Sato2020, Sato2023b, Xiong2018, Chen2022}.

Following the previous considerations, a prototype equation widely used in studies on wave propagation in time-periodic-varying media \cite{Kunz1966,Koutserimpas2018a} is the venerable Mathieu equation \cite{McLachlan1964}. It is among the most well studied equations in physics and has been found to govern the dynamics in many other systems too \cite{Ruby1996}. Typical examples are an inverted pendulum whose pivot point vibrates vertically \cite{Stephenson1908, Polkovnikov2015}, a charged particle in a Paul trap \cite{Paul1990}, a liquid layer that is vertically oscillating \cite{Benjamin1954}, etc. The properties of the Mathieu equation were investigated in numerous classical textbooks \cite{Abramowitz1965,Brillouin1946,Nayfeh1993,Bender1999}, and it is known that both stable and unstable solutions are supported \cite{Teschl2012,Smith2007}. Several experiments have demonstrated the possibility of parametric amplification in platforms that are described by the Mathieu equation (see, for instance, Ref. \cite{Alvaro2021} and the references therein). It has been shown that stable solutions of the Mathieu equation are good candidates to be transiently amplified because the matrix that propagates the initial conditions forward in time  is non-normal \cite{Schmid2001}, yet open questions remain. In particular, in the context of wave propagation in media with harmonically time-modulated propagation speed, which can be mapped to the Mathieu equation, it is natural to ask whether the non-normality-induced transient amplification of the stable Mathieu solutions can be harnessed for a controlled wave amplitude increase.

In this paper we answer this question by first carrying out a comprehensive investigation of the transient amplification of stable Mathieu solutions, corresponding to a wave that propagates in an infinite harmonically time-modulated medium. By an appropriate change of variables, we focus on growth supplementary to evident adiabatic invariance. Owing to the $\epsilon$-pseudospectrum of the monodromy matrix---the matrix that propagates the initial conditions over one period---we reveal that the initial time $t_0$ has a strong impact on the maximum transient amplification. In addition, we provide numerical evidence that the global maximum amplification is captured merely by the monodromy matrix. Then, we consider the case of a wave equation with time interfaces between constant and harmonically modulated propagation speeds. We demonstrate that arbitrary amplification of the wave amplitude can be achieved.

Our work is organized as follows: In Sec. \ref{Section2} we consider the propagation of a wave in an infinite one-dimensional medium that is periodically modulated in time, so that the Mathieu equation emerges. In Sec. \ref{Section3} we briefly review the basic properties of the Mathieu equation, and we derive its stability chart. In Sec. \ref{Section4} we give a few examples of stable solutions that are transiently amplified, and we introduce a measure for the quantification of the transient amplification that filters the one that stems from adiabatic invariance. In Sec. \ref{Section5} we explore the impact of the initial time in these amplifying features, and we explain the underlying physics in terms of the non-normality of the monodromy matrix, while in Sec. \ref{Section6} we calculate the overall maximum amplification of the stable solutions of the Mathieu equation. Then, in Sec. \ref{Section7} we present the evolution of waves (standing and propagating) in the presence of a suitably chosen time interface between a constant and harmonically modulated propagation speed (Floquet medium). We show that a maximum transient amplification is experienced, corresponding to the biggest possible one for the Mathieu solution. Subsequently, by adjusting the number and position of the time interfaces, we demonstrate the achievement of an arbitrary amplification of the wave amplitude. Finally, in Sec. \ref{section9} we summarize our findings.


\section{Wave propagation in a time-varying medium}
\label{Section2}

We follow Ref.~\cite{Koutserimpas2018a}, and we study wave propagation in an infinite harmonically time-modulated medium that is governed by the following wave equation:
\begin{equation}
\frac{\partial^2 \psi (x,\tau)}{ \partial \tau^2} 
=
\left[ \tilde{\delta} - 2 \tilde{q} \cos(\Omega \tau)\right] 
\frac{\partial^2 \psi (x,\tau)}{ \partial x^2},
\label{eqwave}
\end{equation}
where $\tilde{\delta}$, $\tilde{q}$, and $\Omega$ are constants. This wave equation describes the propagation of an electromagnetic wave in a medium with electric permittivity $\epsilon(t)=\epsilon_0/[\tilde{\delta} - 2 \tilde{q} \cos(\Omega \tau)]$ (the speed of light is $c=1$, and $\epsilon_0$ is the vacuum permittivity) \cite{Kunz1966}. It could also correspond to the propagation of an elastic wave in a medium with time-dependent stiffness \cite{Norris2018}. By separation of variables, one class of solutions of Eq.~(\ref{eqwave}) is $\psi(x,\tau)= f(\tau) h(x)$, and by substituting this form into Eq.~(\ref{eqwave}) we arrive at the following set of ordinary differential equations that $f$ and $h$ satisfy:
\begin{gather}
\dfrac{d^2 h(x)}{dx^2}+k^2h(x) =0
\label{eqh}
\\
\dfrac{d^2 f(\tau)}{d\tau^2}+k^2 \left[ \tilde{\delta} - 2 \tilde{q} \cos(\Omega \tau)\right] f(\tau) =0
\label{eqf}
\end{gather}
where $k$ is the real wave number of the wave. From Eq.~(\ref{eqh}) we find that $h(x)$ has the form $h(x) \sim e^{\pm i k x}$, while Eq.~(\ref{eqf}), after time rescaling $t=\Omega \tau/2$ and setting $\delta=4 k^2 \tilde{\delta}/\Omega^2$, and $q=4 k^2 \tilde{q}/\Omega^2$, drops to the usual form of the Mathieu equation, that is,
\begin{equation}
\label{eqMathieu}
\ddot{f}+
\omega^2(t)
 f=0
\end{equation}
with $\omega^2(t)=\delta -2q \cos(2t)$. Note that the overdots represent differentiation with respect to the time $t$. The Mathieu equation contains both stable and unstable---exponentially growing---solutions, according to the values of the parameters $(\delta,q)$. The amplification is usually related to the exponentially growing solutions, namely, to the parametric instability. However, a wave can experience amplification even with asymptotic stability \cite{Trefethen2005}.  In this case the amplification is a transient phenomenon characterizing the stable solutions of the Mathieu equation. We will perform a detailed analysis of this transient amplification employing suitable tools for its quantitative description.
Before proceeding with this analysis we briefly review the Mathieu equation.


\section{Review of the Mathieu equation}
\label{Section3}

The Mathieu equation written as a system of two linear first order differential equations has the form
\begin{equation}
\label{eqMathieu2}
\dot{ {\boldsymbol{\eta}}}(t) = {\mathbf{A}}(t) {\boldsymbol{\eta}}(t),
\end{equation}
with 
$
\boldsymbol{\eta}(t)
=
\begin{pmatrix}
f(t) \\
\dot{f}(t)
\end{pmatrix}
$
and 
$
{\mathbf{A}}(t)
=
\begin{pmatrix}
0 && 1 \\
- \omega^2(t) && 0
\end{pmatrix}
$.
The general solution of Eq.~(\ref{eqMathieu2}) can be written in the form
\begin{equation}
\label{solMathieu}
{\boldsymbol{\eta}}(t)=
\mathbf{\Psi}(t,t_0) {\boldsymbol{\eta}}(t_0),
\end{equation}
with ${\boldsymbol{\eta}}(t_0)$ being the initial condition.
The matrix 
$
\mathbf{\Psi}(t,t_0)
$, which evolves the initial vector in time, will be called the principal matrix solution \cite{Teschl2012}.  

\begin{figure}
\begin{center}
\includegraphics[width=1\columnwidth]{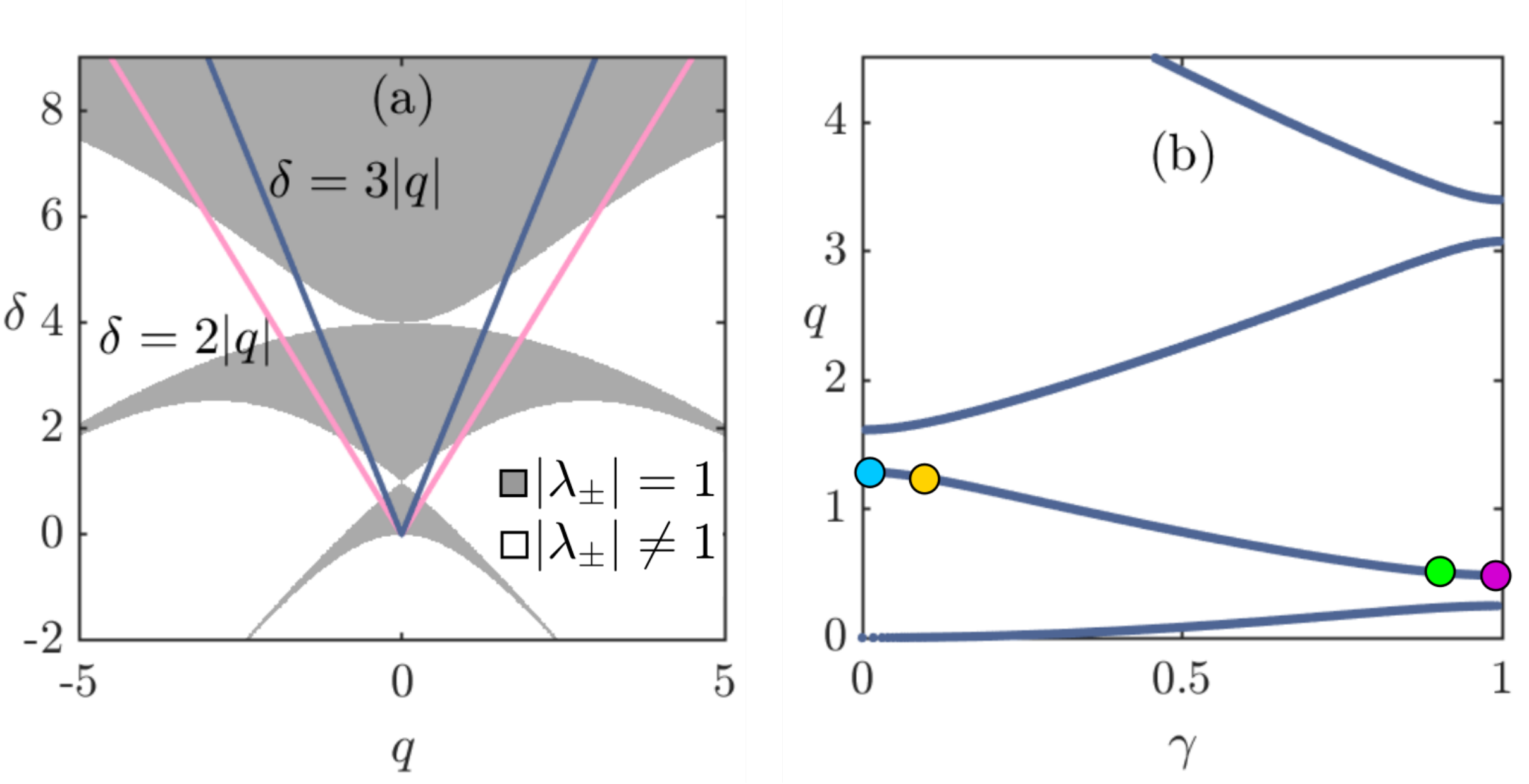}
\caption
{
Stability diagrams of the Mathieu equation (\ref{eqMathieu}).
(a) The norm of the eigenvalues $\lambda_{\pm}$ of the monodromy matrix
as a function of parameters $\delta$ and $q$. 
Stable regions are in gray.
Also shown are two cuts of the form $\delta=c|q|$, with $c=2,3$. 
(b) Floquet exponent $\gamma$ as the line $\delta=3|q|$ is scanned. 
The circles correspond to the four cases displayed in Fig.~\ref{fig2}.
}
\label{fig1}
\end{center}
\end{figure}

The matrix $\mathbf{A}(t)$ that contains the parameters of the Mathieu equation is $\pi$ periodic, i.e., $\mathbf{A}(t)=\mathbf{A}(t+\pi)$. Therefore, Floquet theory applies and states that the stability properties of the solutions can be deduced from the eigenvalues of the matrix ${\mathbf{\Psi}}(t_0+\pi,t_0)$, called the monodromy matrix. These eigenvalues, which we denote as ${\lambda}_{\pm}$ and which are commonly called Floquet multipliers, do not depend on the choice of the initial time $t_0$ since the matrices ${\mathbf{\Psi}}(\pi+t_1,t_1)$ and ${\mathbf{\Psi}}(\pi+t_2,t_2)$ are similar \cite{Teschl2012}. 

\begin{figure*}
\begin{center}
\includegraphics[width=1.4\columnwidth]{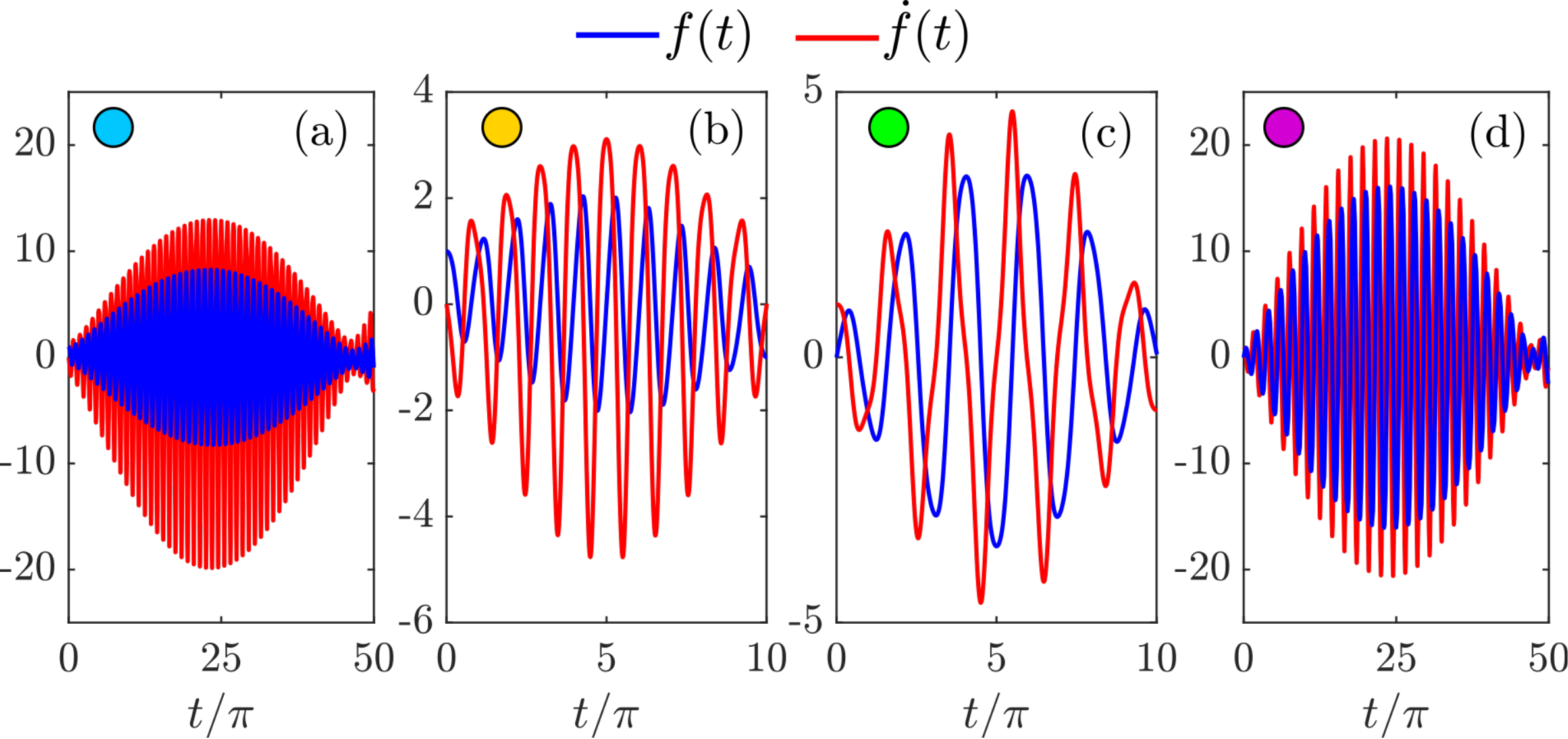}
\caption
{
Typical solutions of the Mathieu equation (\ref{eqMathieu}) in the stable regime.
In all cases the sets of parameters $(\delta,q)$ lie in the line $\delta=3q$.
(a) Evolution of the initial conditions $f(0)=1$ and $\dot{f}(0)=0$ when $q = 1.285$ and $\gamma = 0.02$. 
This set of parameters is indicated by the blue circle in Fig.~\ref{fig1}(b). 
(b) Same as (a), but $q = 1.239$ and $\gamma  = 0.1$  [yellow circle in Fig.~\ref{fig1}(b)]. 
(c) Same as (a), but $q = 0.507$ and $\gamma = 0.9$ [green circle in Fig.~\ref{fig1}(b)].
Also, the initial conditions are $f(0)=0$ and $\dot{f}(0)=1$ in this case.
(d) Same as (c), but $q = 0.4855$ and $\gamma = 0.98$ [purple circle in Fig.~\ref{fig1}(b)].
}
\label{fig2}
\end{center}
\end{figure*}

From Liouville's formula, $\det \left[ {\mathbf{\Psi}}(t_0+\pi,t_0) \right] = \exp\left[ \int_{t_0}^{t_0+\pi} \text{Tr}\mathbf{A}(s) ds \right]$, it follows that the determinant of the monodromy matrix ${\mathbf{\Psi}}(t_0+\pi,t_0)$ is 1, and therefore, its two eigenvalues $\lambda_{\pm}$ satisfy the relation $\lambda_+ \lambda_- = 1$. When $|\lambda_{\pm}|=1$, they are complex conjugates and are restricted to lie in the unit circle in the complex plane: The solutions are stable. When $|{\lambda}_{\pm}| \neq 1$, the typical solutions are unstable and grow exponentially with time. Figure~\ref{fig1}(a) illustrates the norm of these eigenvalues for each pair of the only parameters in Eq. (\ref{eqMathieu}) ($\delta$ and $q$). The gray region in this chart (apart from the boundaries) corresponds to stable solutions, while in the white region $|{\lambda}_{\pm}| \neq 1$. The boundary between these two regions corresponds to exceptional points where the typical solutions grow linearly with time. This plot is widely known as the stability chart of the Mathieu equation \cite{McLachlan1964}.

The stability properties of periodic systems are usually studied in terms of the Floquet exponent $\gamma$, which is related to the Floquet multipliers by $\lambda_{\pm}=e^{\pm i \gamma \pi}$. In Fig.~\ref{fig1}(b) we present the exponent $\gamma$ along the cut $\delta=3q$ ($\delta/q$ is constant for constant $\tilde{\delta}$, $\tilde{q}$, and $\Omega$). 
This is a Floquet spectrum, with the bands corresponding to the stable regions and the gaps corresponding to the unstable ones \cite{Koutserimpas2018}.


\section{Transient amplification}
\label{Section4}

In this section and in the remainder of this work, we will elaborate on the transient amplification that is displayed by the stable solutions of the Mathieu equation (Fig.~\ref{fig1}). To illustrate this, in Figs.~\ref{fig2}(a)-\ref{fig2}(d) we plot $f(t)$ and $\dot{f}(t)$ for the four cases  that are shown in Fig.~\ref{fig1}(b). In Figs.~\ref{fig2}(a) and \ref{fig2}(a)(b) the initial conditions are $f(0)=1$ and $\dot{f}(0)=0$, while in Figs.~\ref{fig2}(c) and \ref{fig2}(d) the corresponding initial conditions are $f(0)=0$ and $\dot{f}(0)=1$. Note that in all cases the closer to the edges of the bands are, the stronger the amplification is. In addition, we observe that the transient amplification time interval increases without limit as we approach the unstable region. This characteristic behavior resembles the scale-free localization of the critical non-Hermitian skin effect \cite{Murakami2021, Gong2020}, where the localization length increases with increasing size of the system, becoming infinite at the critical point. In our case, it is the amplification time interval that diverges as the distance to the unstable region decreases. This transient amplification cannot be captured by the stability analysis since the Floquet exponents are purely imaginary in all these examples.
This is due to the non-normality of the  principal matrix $\mathbf{\Psi}(t,t_0)$ \cite{Trefethen2005,commentnonlinear}.


\subsection{Choice of variables}

We choose to change the variables to the following ones:
\begin{equation}
\label{transf}
X=f\sqrt{\omega(t)},~~~Y=\dot{f}/\sqrt{\omega(t)}.
\end{equation}
To illuminate the utility of this transformation one should consider the Wentzel-Kramers-Brillouin (WKB) limit of Eq.~(\ref{eqMathieu}), which is when $\omega(t)$ varies slowly with time, i.e., $\Omega \ll \omega$.
We can show then that in the WKB limit the norm of the vector 
$
{\boldsymbol{\xi}}(t)
=
\begin{pmatrix}
X(t)
\\
Y(t)
\end{pmatrix}
$,
i.e., $||\boldsymbol{\xi}(t)||=\sqrt{|X(t)|^2+|Y(t)|^2}$, is constant and equal to $\sqrt{|X(0)|^2+|Y(0)|^2}$ (it is the adiabatic invariant of Eq.~(\ref{eqMathieu}) \cite{Bliokh2003}). This WKB adiabatic invariant already predicts amplification with the WKB solution given by $f(t)=e^{\pm i\int_0^t\omega(s)ds} / \sqrt{\omega(t)}$, but in this paper we will try to go beyond this adiabatic effect, and we filter this by choosing these new variables \cite{comment1}. Away from the WKB limit the norm of the vector ${\boldsymbol{\xi}}(t)$ is not constant: Nontrivial amplification is captured, nontrivial in the sense that it is not predicted by WKB. In Appendix \ref{AppendixA} we present some examples showing the convergence to the WKB limit for large parameters $\delta$ and $q$.

Using Eqs.~(\ref{eqMathieu2}) and (\ref{transf}), we get
\begin{equation}
\dot{ \boldsymbol{\xi}}(t) = \mathbf{C}(t) \boldsymbol{\xi}(t),
\end{equation}
where $\mathbf{C}(t)$ is the $\pi$-periodic matrix
$
\begin{pmatrix}
\dot{\omega}/2\omega && \omega \\
-\omega && -\dot{\omega}/2\omega
\end{pmatrix}
$.
Moreover, the vector $\boldsymbol{\xi}(t)$ is given in terms of the initial state vector $\boldsymbol{\xi}(t_0)$ as
\begin{equation}
{\boldsymbol{\xi}}(t)=
\mathbf{\Phi}(t,t_0) {\boldsymbol{\xi}}(t_0),
\end{equation}
where the matrix $\mathbf{\Phi}(t,t_0)$ is expressed in terms of the principal matrix ${\mathbf{\Psi}}(t,t_0)$ through the relation 
\begin{equation}
{\mathbf{\Phi}}(t,t_0)
=
\begin{pmatrix}
\dfrac{\sqrt{\omega(t)}}{\sqrt{\omega(t_0)}} {\Psi}_{11}(t,t_0) && \sqrt{\omega(t)\omega(t_0)} {\Psi}_{12}(t,t_0) \\
\dfrac{1}{\sqrt{\omega(t)\omega(t_0)}}{\Psi}_{21}(t,t_0) && \dfrac{\sqrt{\omega(t_0)}}{\sqrt{\omega(t)}} {\Psi}_{22}(t,t_0)
\end{pmatrix}.
\end{equation}
Since $\omega(t_0+\pi)=\omega(t_0)$, the monodromy matrices ${\mathbf{\Phi}}(t_0+\pi,t_0)$ and $\mathbf{\Psi}(t_0+\pi,t_0)$ have the same eigenvalues, namely, the Floquet multipliers $\lambda_{\pm}$.


\subsection{Choice of measure}

As we noted before, to avoid amplification simply obtained with adiabatic invariance, we will use the norm of $\boldsymbol{\xi}(t)$ as a measure of the amplification. In particular, the maximum possible amplification is given by the maximum of the norm of $\boldsymbol{\xi}(t)$ at a given $t$ over all the initial conditions $\boldsymbol{\xi}(t_0)$ at a given $t_0$. This is equivalent to the 2-norm  of the matrix ${\mathbf{\Phi}}(t,t_0)$ since, by definition, this matrix norm is given by
\begin{equation}
||\mathbf{\Phi}(t,t_0)||=\max_{\boldsymbol{\xi}(t_0),||\boldsymbol{\xi}(t_0)||=1} ||\mathbf{\Phi}(t,t_0) \boldsymbol{\xi}(t_0)||
.
\end{equation}
Therefore, the quantity $||\mathbf{\Phi}(t,t_0)||$ reveals the maximum possible amplification of the vector ${\boldsymbol{\xi}}(t)$ at time $t$, out of all the initial conditions at $t_0$.

The norm of $\mathbf{\Phi}(t,t_0)$ and the corresponding maximizing initial condition ${\boldsymbol{\xi}}(t_0)$ are provided by the singular value decomposition (SVD) \cite{Steward1993}. The SVD of a real matrix is the decomposition $\mathbf{\Phi}(t,t_0) = \mathbf{U}(t,t_0) \mathbf{\Sigma}(t,t_0) \mathbf{V}^{T}(t,t_0)$, where  $\mathbf{\Sigma}(t,t_0)$ is a diagonal matrix with real and non-negative entries that are arranged in descending order. Also, $\mathbf{U}(t,t_0)$ and $\mathbf{V}(t,t_0)$ are orthogonal matrices, and $T$ denotes the transpose. The largest singular value $\sigma_{max}(t,t_0)$ [which is the first element of $\mathbf{\Sigma}(t,t_0)$]  is the norm $||\mathbf{\Phi}(t,t_0)||$. Furthermore, the SVD also provides the most amplified initial condition ${\boldsymbol{\xi}}(t_0)$: the first column of the matrix $\mathbf{V}(t,t_0)$.

\begin{figure}
\begin{center}
\includegraphics[width=1\columnwidth]{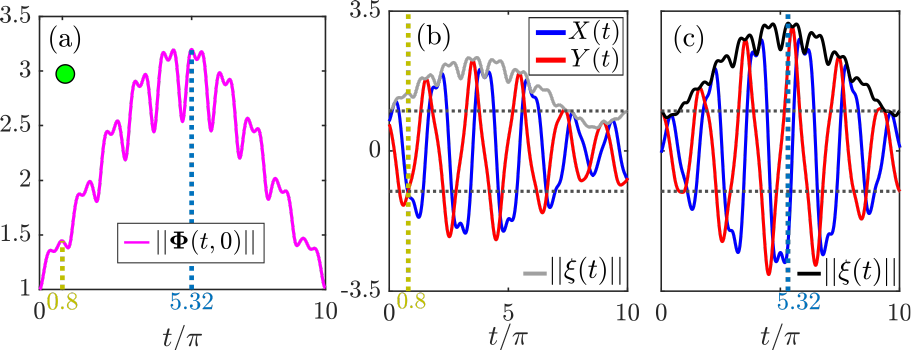}
\caption
{
Evolution of the norm of the propagator. 
The set of parameters $(\delta,q)$ that is used here is the one used in Fig.~\ref{fig2}(c), namely, $q = 0.507$ and $\delta=3q$, which result in a Floquet exponent $\gamma=0.9$.
(a) The norm of the propagator matrix $\mathbf{\Phi}(t,0)$.
(b) Evolution of the initial conditions $X(0)=0.7469$ and $Y(0)=0.6649$, which yield the maximum norm of the vector $\boldsymbol{\xi}(t)$ at time $t=0.8\pi$. 
The solid curve is the norm of $\boldsymbol{\xi}(t)$. 
(c)  Same as (b), with initial conditions $X(0)=-0.0449$ and $Y(0)=0.999$, which yield the maximum norm of the vector $\boldsymbol{\xi}(t)$ at time $t=5.32\pi$. 
Also shown by a black solid curve is the norm of $\boldsymbol{\xi}(t)$.
In (b) and (c), the dotted lines represent the quantities $\pm \sqrt{|X(0)|^2+|Y(0)|^2}=\pm 1$.
}
\label{fig3}
\end{center}
\end{figure}

Figure~\ref{fig3}(a) illustrates the norm of the propagator $\mathbf{\Phi}(t,0)$ as a function of the time $t$ for the same set of parameters $(\delta,q)$ as used in Fig.~\ref{fig2}(c). The norm clearly  exceeds 1, showing the existence of transient amplification in the stable regime. Moreover, the norm of the propagator is periodic  when  the exponent $\gamma=m_1/m_2$ and has a period of at most $m_2 \pi$ \cite{Abramowitz1965}. Therefore, in this example where $\gamma = 0.9$, the norm of the propagator oscillates with a period of $10 \pi$. In addition, the norm of $\mathbf{\Phi}(t,0)$ at time $t=5.32 \pi$ is the maximum possible amplification that we can get  for this set of parameters with $t_0=0$. In Figs.~\ref{fig3}(b) and \ref{fig3}(c) we present the evolution with time of the variables $X$ and $Y$ when two different initial conditions are considered. Both of these initial conditions yield the maximum value of $||\boldsymbol{\xi}(t)||$, but at two different "final" times $t$, $t=0.8\pi$ in \ref{fig3}(b) and $t=5.32\pi$ in \ref{fig3}(c). We present in Figs.~\ref{fig3}(b) and \ref{fig3}(c) the corresponding norms of $\boldsymbol{\xi}$. Also shown are the quantities $\pm \sqrt{|X(0)|^2+|Y(0)|^2}=\pm 1$ (adiabatic prediction) in order to clearly see the nontrivial amplification not predicted by simple adiabatic invariance. We note here that similar transient amplifying phenomena occur for other time-dependent systems as well [see Appendix \ref{AppendixB} for the Meissner equation with piecewise constant frequency $\omega(t)$].

\subsection{Floquet representation and pseudospectrum}

\begin{figure}
\begin{center}
\includegraphics[width=1\columnwidth]{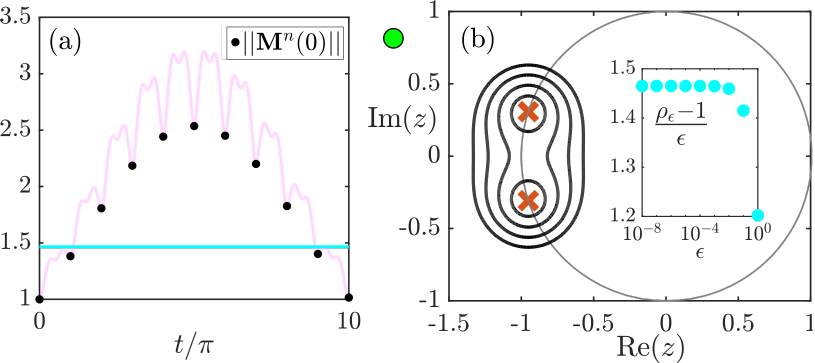}
\caption
{
Amplification in terms of the monodromy matrix and its $\epsilon$-pseudospectrum.
We use the same set of parameters $(\delta,q)$ as in Fig.~\ref{fig3}.
(a) 
The black dots denote the quantity $||{\mathbf{M}^n(0)}||$, and the magenta line denotes the norm $||{\mathbf{\Phi}(t,0)}||$.
The cyan line corresponds to the Kreiss constant, which provides a bound for ${\max_n ||{\mathbf{M}^n(0)}||}$. 
(b) Crosses show the eigenvalues of $\mathbf{M}(0)$ which lie in the unit circle (shown by the solid gray line). 
The black solid lines are the boundaries of the $\epsilon$-pseudospectrum for $\epsilon=$0.08, 0.14, 0.2, 0.26.
The inset shows the quantity on the right-hand side of Eq.~(\ref{eqbound}), which is a bound for the amplification. 
The shown plateau corresponds to the Kreiss constant, displayed in (a) by the cyan line.
}
\label{fig4}
\end{center}
\end{figure}

In this part, to describe the amplification we will focus on the monodromy matrix.
The Floquet theory states that the propagator ${\mathbf{\Phi}}(t,t_0)$ is written in the form \cite{Teschl2012}
\begin{equation}
{\mathbf{\Phi}}(t,t_0)={\mathbf{P}}(t,t_0) e^{\mathbf{B}(t_0)(t-t_0)} ,
\end{equation}
where the matrix ${\mathbf{P}}(t,t_0)$ is $\pi$ periodic at both times $t$ and $t_0$, while the matrix $\mathbf{B}(t_0)$ depends only on the initial time $t_0$.
The monodromy matrix will be denoted by $\mathbf{M}(t_0)$ as
\begin{equation}
\mathbf{M}(t_0)=e^{\mathbf{B}(t_0)\pi}.
\label{e	monodromy}
\end{equation}
Iterated powers $||\mathbf{M}^n(t_0)||$ equal to $||{\mathbf{\Phi}(t_0+n\pi,t_0)}||$ provide the maximum possible amplification at each multiple of $\pi$ and a stroboscopic view of the amplification.
This is illustrated in Fig.~\ref{fig4}(a), where we present $||\mathbf{M}^n(0)||$ and the associated norm of the propagator $||\mathbf{\Phi}(t,0)||$ for the same set of parameters as in Fig.~\ref{fig3}. 
We already see that this stroboscopic point of view gives useful hints about the amplification.
We now concentrate on finding the lower bound for ${ \max_n ||{\mathbf{M}^n(t_0)}||}$ using the concept of the pseudospectrum.

The $\epsilon$-pseudospectrum \cite{Trefethen2005} of the matrix $\mathbf{M}(t_0)$ is defined as the set of all complex numbers $z$ such that 
\begin{equation}
||[z-\mathbf{M}(t_0)]^{-1}||>\epsilon^{-1},
\end{equation}
with $\epsilon>0$.
Note that the eigenvalues are points corresponding to $\epsilon \rightarrow 0$.
Figure~\ref{fig4}(b) shows the boundaries of the $\epsilon$-pseudospectrum of $\mathbf{M}(0)$ for different values of $\epsilon$ where the eigenvalues of $\mathbf{M}(0)$ appear as singularities; the behavior of the pseudospectrum around these singularities  directly gives the lower bound for amplification.
Indeed, the pseudospectrum provides several useful bounds.
For instance, the maximum value of $||{\mathbf{M}^n(t_0)}||$ can be estimated by
\begin{equation}
\label{eqbound}
\max_{n} ||\mathbf{M}^n(t_0)|| \geq \max_{\epsilon} \dfrac{\rho_{\epsilon}(\mathbf{M}(t_0))-1}{\epsilon}
\end{equation}
where $\rho_{\epsilon}(\mathbf{M}(t_0))$ is the so called $\epsilon$-pseudospectrum radius, given by
\begin{multline}
\rho_{\epsilon}(\mathbf{M}(t_0))= \\
\max
\left\lbrace
|z|:z \in \mathbb{C}, ||(z-\mathbf{M}(t_0))^{-1}||>\epsilon^{-1}
 \right\rbrace 
 .
\end{multline}
The quantity on the right-hand side of Eq.~(\ref{eqbound}) is the Kreiss constant \cite{Mitchell2020}. 
If the  Kreiss constant is more than 1, then non trivial amplification is captured. 
In the inset of Fig.~\ref{fig4}(b) we present the quantity $[\rho_{\epsilon}(\mathbf{M}(0))-1]/\epsilon$ as a function of $\epsilon$. 
The plateau that is shown determines the Kreiss constant, which exceeds 1, indicating amplification [see also Fig.~\ref{fig4}(a)].

\section{Impact of the initial time}
\label{Section5}

\begin{figure}
\begin{center}
\includegraphics[width=1\columnwidth]{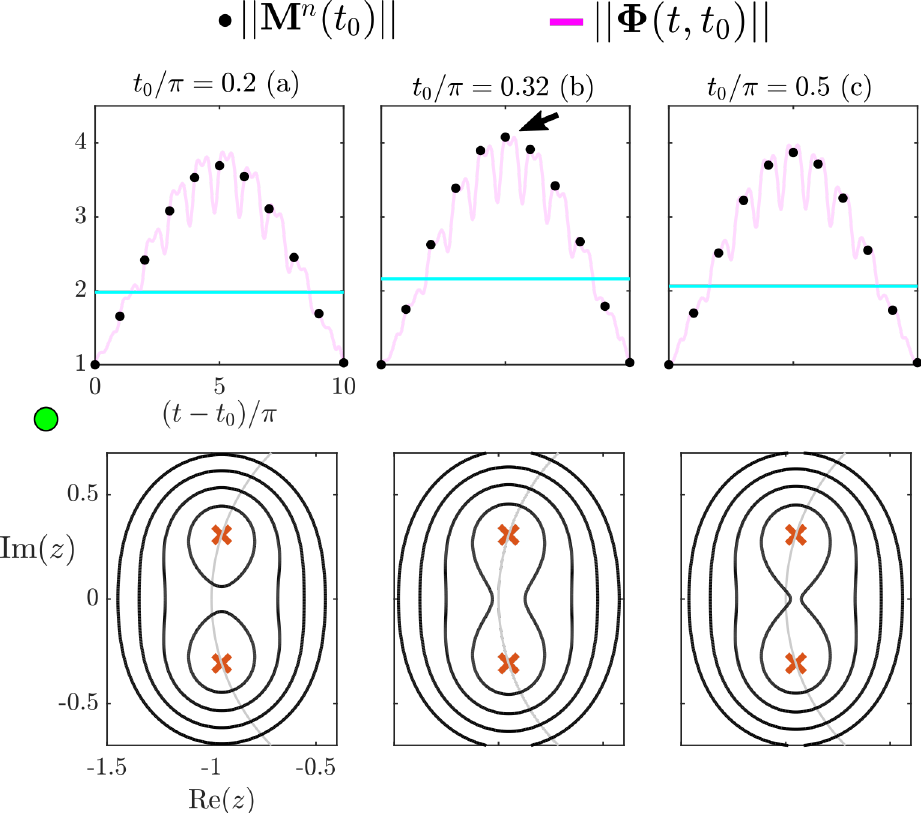}
\caption
{
Influence of the initial time $t_0$.
In all cases we use the same set of parameters $(\delta,q)$ as in Fig.~\ref{fig3}.
Top: The quantities $||\mathbf{\Phi}(t,t_0)||$ as a function of $t$ (magenta lines) and $||\mathbf{M}^n(t_0)||$ as a function of $n=0,1,...,10$ (black dots) for three initial times: (a) $t_0=0.2$, (b) $t_0=0.32$, and (c) $t_0=0.5$. 
Also shown by the cyan lines are the  Kreiss constants. 
Bottom: Corresponding boundaries of the pseudospectra for $\epsilon=$0.08, 0.14, 0.2 and 0.26. Also shown by the crosses are the Floquet multipliers, which lie in the unit circle and do not change as $t_0$ changes.
}
\label{fig5}
\end{center}
\end{figure}

Until now, $t_0$ was assumed to be zero.
\textit{A priori}, there is no reason that it gives the best amplification; thus, we are now going to investigate other values of the initial time.
In the top three panels in Figs.~\ref{fig5}(a)-\ref{fig5}(c) we show the norm $||\mathbf{\Phi}(t,t_0)||$  as a function of $t$ for three different choices of $t_0$.
In the same panels we also show the stroboscopic monodromy norm $||\mathbf{M}^n(t_0)||$ as a function of $n$.
These panels indicate that the initial time $t_0$ has an influence that should be taken into account. 
Going into the evaluation of the amplification lower bound, we face an interesting situation:
With varying $t_0$ the pseudospectrum of $\mathbf{M}(t_0)$ evolves, but its eigenvalues are pinned at fixed positions
[due to the similarity of two matrices, $\mathbf{M}(t_1)$ and $\mathbf{M}(t_2)$; see Sec. \ref{Section3}].
This is illustrated in Fig.~\ref{fig5} (bottom panels), which report the evolution of the pseudospectrum (as well as the non-normality) of $\mathbf{M}(t_0)$.
This evolution with $t_0$ is also reflected by the change in the lower bound given by the Kreiss constant.

\begin{figure}
\begin{center}
\includegraphics[width=1\columnwidth]{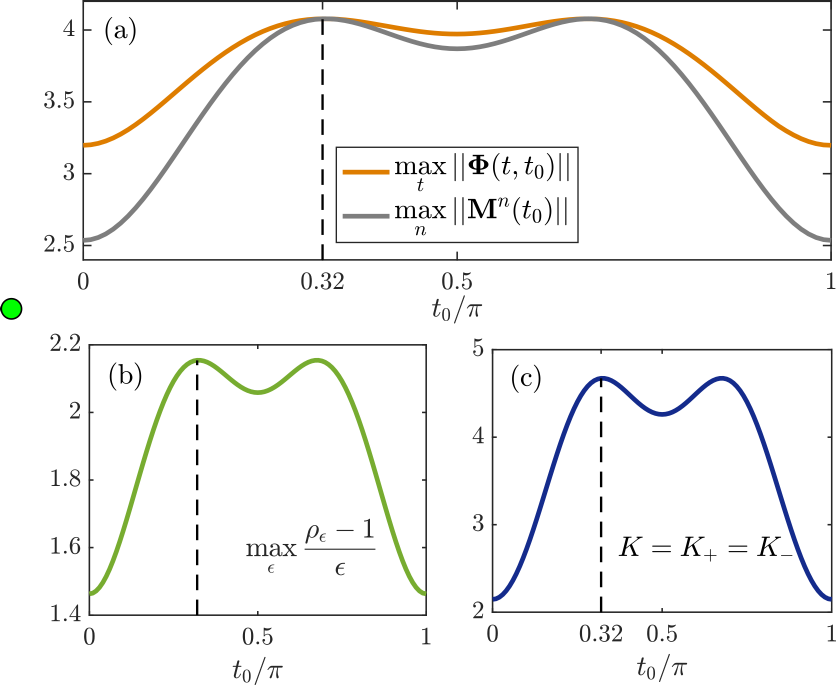}
\caption
{
Evolution with $t_0$ of the nonnormality of the propagator.
We use the same set of parameters $(\delta,q)$ as in Fig.~\ref{fig3}.
(a) ${\max_t ||\mathbf{\Phi}(t,t_0)||}$ and the quantity ${\max_n ||\mathbf{M}^n(t_0)||}$, where $n$ is an integer number.
(b) Kreiss constant, i.e., ${\max_{\epsilon} \lbrace [\rho_{\epsilon}(\mathbf{M}(t_0))-1]/\epsilon} \rbrace$, as a function of the initial time $t_0$.
(c) The Petermann factors $K=K_{+}=K_{-}$ of the monodromy matrix as the initial time $t_0$ changes. 
The Petermann factors measure the parallelism of its right and left eigenvectors.
}
\label{fig6}
\end{center}
\end{figure}

A closer look at Fig.~\ref{fig5} reveals that for $t_0=0.32 \pi$, the maximum amplification is provided merely by the monodromy matrix. 
This observation drives us to investigate whether something special happens for this particular initial time.
To that end,  in Fig.~\ref{fig6}(a) we illustrate the quantities ${\max_t ||\mathbf{\Phi}(t,t_0)||}$ \cite{comment2} and ${\max_n ||\mathbf{M}^n(t_0)||}$ as a function of $t_0$, and we observe that their maxima coincide for $t_0=0.32 \pi$.
The overall maximum amplification is captured merely by the monodromy matrix.
More insight into this last result is provided by studying the Kreiss constant for all $t_0$.
It displays the same pattern as before, with its maximum for $t_0=0.32 \pi$ [see Fig.~\ref{fig6}(b)].
Let us note here that for fixed $t_0$ the norm of the propagator $\max_t ||\mathbf{\Phi}(t)||$ exhibits a power law increase as the system moves closer to the instability region, in the form $\max_t ||\mathbf{\Phi}(t)|| \sim |q-q^*|^{-1/2}$, where $q^*$ is the closest point on the instability boundary.

Another measure that examines the non-normality of a matrix is its Petermann factors \cite{Trefethen2005,Berry2003,Lee2009} (or conditioning number).
The two Petermann factors of the monodromy matrix are given by
\begin{equation}
K_{\pm}
=
\dfrac{||u_{\pm}|| ~ ||v_{\pm}||}{|v^{\dagger}_{\pm} u_{\pm}|},
\end{equation}
where $\boldsymbol{u}_{\pm}(t_0)$ and $\boldsymbol{v}_{\pm}(t_0)$ correspond, respectively, to right and left eigenvectors associated with eigenvalues $\lambda_{\pm}$.
For a normal matrix the Petermann factors are equal to 1.
In the case of the monodromy matrix  in our problem the two Petermann factors $K_{\pm}$ are equal because its eigenvalues are complex conjugates in the stable region and therefore the two right and left eigenvectors 
are also complex conjugates, namely, $\boldsymbol{u}_+=\overline{\boldsymbol{u}}_-$ and $\boldsymbol{v}_+=\overline{\boldsymbol{v}}_-$.
In Fig.~\ref{fig6}(c) we present the Petermann factor $K=K_{+}=K_{-}$ of the monodromy matrix as a function of the initial time  for the same set of parameters $\delta$ and $q$ as used in Figs.~\ref{fig6}(a) and \ref{fig6}(b).
Figure~\ref{fig6}(c) confirms that non-normality is maximum for $t_0=0.32\pi$.

From the analysis of this section, it appears that the monodromy matrix is capable of determining the overall maximum amplification for the Mathieu equation.
The latter occurs when the non-normality of the monodromy matrix is maximal.
This property seems to be general, as verified through numerical investigations for a dense set of parameters $\delta$ and $q$.


\section{Maximum transient amplification: Monodromy matrix description}
\label{Section6}

\begin{figure}
\begin{center}
\includegraphics[width=1\columnwidth]{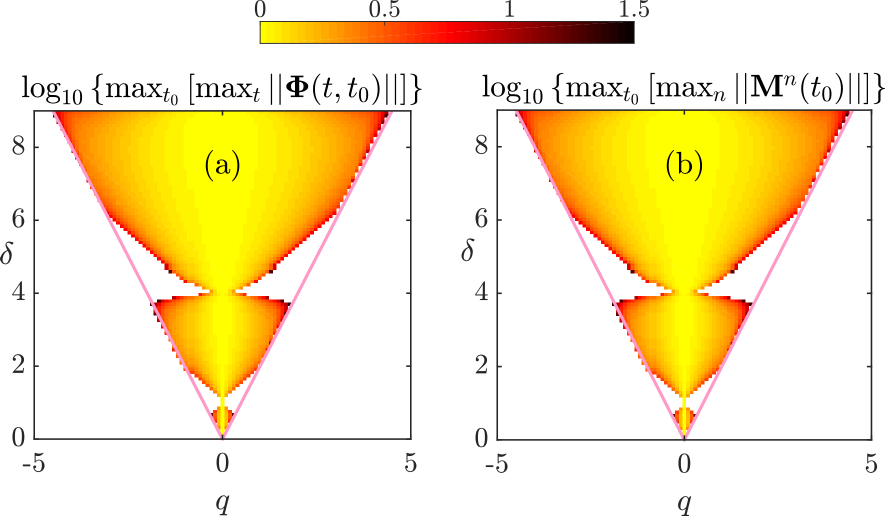}
\caption
{
Global maximum amplification over all initial conditions and all initial times.
(a) The quantity  ${\log_{10} } \left\{ \max_{t_0} \left[ \max_t ||\mathbf{\Phi}(t,t_0)|| \right] \right\}$ 
and (b) ${ \log_{10} } \left\{  \max_{t_0} \left[ \max_n ||\mathbf{M}^n(t_0)|| \right]  \right\} $.
These quantities are calculated at the stable regime inside the cone that is formed by the line $\delta=2|q|$.
}
\label{fig7}
\end{center}
\end{figure}

The goal of this section is to calculate the maximum possible amplification of all the stable solutions. 
In this section we study the quantity
\begin{equation}
\label{eqmax}
\max_{t_0} \left[ \max_t ||\mathbf{\Phi}(t,t_0)|| \right],
\end{equation}
in the stable region of the stability chart \cite{comment3}. 
Figure~\ref{fig7}(a) displays the global maximum [Eq.~(\ref{eqmax})] in the parameter plane $(\delta,q)$.
Note that we exclusively consider  the case of positive $\omega^2$ or positive permittivity, restricting our analysis to the parameter domain inside the cone $\delta=2|q|$.
It is clear that the solutions close to the unstable region are intensively amplified. 
In fact, this maximum amplification diverges as the boundary with the unstable region is approached.
Obviously, along the line $\delta=0$ no amplification is captured since the Mathieu equation drops to the equation of the harmonic oscillator.
The comparison with the stroboscopic monodromy norm is displayed in Fig.~\ref{fig7}(b) with  
$
{
\max_{t_0} \left[ \max_n ||\mathbf{M}^n(t_0)|| \right]
}.
\label{eqmax2}
$
Comparing Figs.~\ref{fig7}(a) and \ref{fig7}(b) confirms that the monodromy matrix is able to predict the amplification.
These results support our conjecture that the monodromy matrix determines the overall maximum amplification exhibited by the stable solutions of the Mathieu equation.

\section{Back to the wave propagation and physical implications}
\label{Section7}

\subsection{Wave propagation in Mathieu media}
In this section we give an example of wave evolution 
given by $\psi(kx,t)=h(kx)f(t)$, with $h(kx)=e^{ikx}$, where $f$ is a solution of the Mathieu equation (\ref{eqMathieu}).
Choosing fixed but arbitrary values of $k$ and $\Omega$, $\psi$ becomes a solution of Eq.~(\ref{eqwave}). 

\begin{figure}
\begin{center}
\includegraphics[width=1\columnwidth]{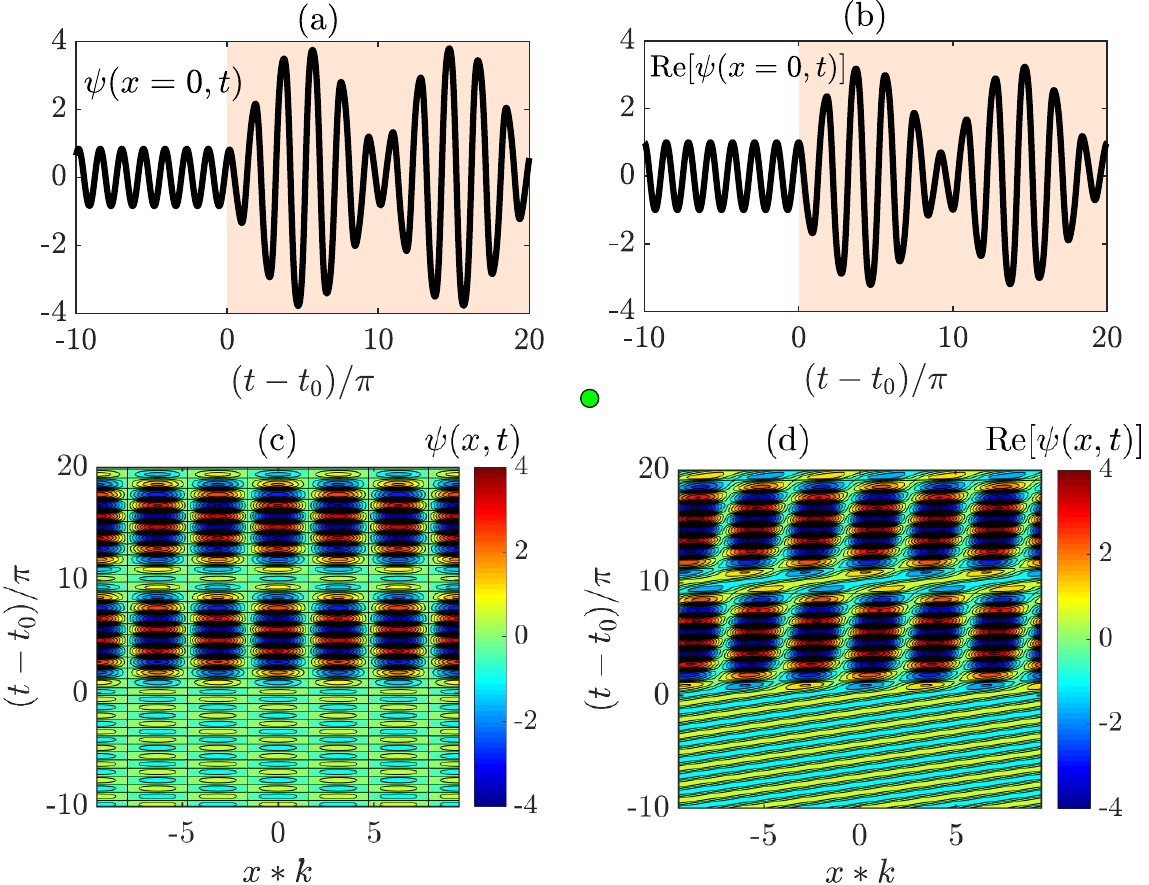}
\caption
{
Space-time evolution of $\psi(x,t)$ satisfying Eq.~(\ref{eqwave}).
The initial time is $t_0=0.32\pi$.
For $t > t_0$ ($t < t_0 $), shown by the orange background (white background),  the wave propagates in a medium with time-varying (constant) permittivity.
(a) We use as initial conditions at $t = t_0$  the optimally amplified ones provided by the SVD. 
(b) We consider a right-going traveling wave for $t < t_0$ with $\psi(kx,t)=e^{i[kx-\omega (t-t_0)]}$ and $\omega^2=3q-2q\cos(2t_0)$. The value of $q = 0.507$ is the same as in Fig.~\ref{fig2} (green point).
(c) Spatiotemporal evolution corresponding to (a).
(d) Spatiotemporal evolution corresponding to (b).
}
\label{fig8}
\end{center}
\end{figure}

We consider in Fig.~\ref{fig8} the wave evolution of a standing wave [Figs.~\ref{fig8}(a) and \ref{fig8}(c)] and a traveling wave [Figs.~\ref{fig8}(b) and \ref{fig8}(d)] with a time-dependent frequency of the form
\begin{equation}
\omega^2(t)
=
\begin{cases}
\delta-2q\cos(2 t_0)=\omega_1^2, & \text{for}~ t < t_0\\
\delta-2q\cos(2 t)=\omega_2^2(t), & \text{for}~ t \geq t_0,
\end{cases}
\label{omega2t}
\end{equation}
where $(\delta,q)$ are the same as in Fig.~\ref{fig5}(b). Note that $t_0$, defining the time interface position,  maximizes the norm in Eq.~\eqref{eqmax} for these values of $(\delta,q)$.
Since $\omega_2^2(t_0)=\omega_1^2$, the frequency $\omega(t)$ is continuous at $t_0$. In Fig.~\ref{fig8}(a) we show the time evolution of the solution at $x=0$, while in Fig.~\ref{fig8}(c) we show the entire spatiotemporal profile of the standing wave solution.

The solution $\psi(kx,t)$, with $\omega(t)$ given by Eq.~(\ref{omega2t}), becomes 
\begin{equation}
\psi(kx,t)=\begin{cases}
f_1(t) h(kx) = f_1(t)e^{ikx}, & \text{for}~ t < t_0,\\
f_2(t) h(kx) = f_2(t)e^{ikx}, & \text{for}~ t \geq t_0,
\end{cases}
\end{equation}
where $f_1(x)$ satisfies the harmonic oscillator equation with frequency $\omega_1$, i.e., $\ddot{f}_1+\omega_1^2 f_1=0$, while $f_2(x)$ satisfies the Mathieu equation with frequency $\omega_2^2(t)$, i.e., $\ddot{f}_2+\omega_2^2(t) f_2=0$. 
The temporal part $f_1(t)$ of the solution is given by
\begin{equation}
f_1(t)=x_0 \cos[\omega_1(t-t_0)] + \frac{y_0}{\omega_1}\sin[\omega_1(t-t_0)],
\end{equation}
where $x_0$ and $y_0$ are the initial conditions. The function $f_2(t)$ is computed numerically with initial conditions 
$f_2(t_0)=f_1(t_0)=x_0$ and $\dot{f_2}(t_0)=\dot{f_1}(t_0)=y_0$.

The standing wave optimal amplification shown in Figs.~\ref{fig8}(a) and \ref{fig8}(c), 
is obtained by maximizing the propagator norm using SVD in the canonical variables defined in Eq.~\eqref{transf}
at times $T=n(5\pi+t_0)$ ($n=1,2,...$). This maximization procedure leads to $t_0=0.32 \pi$, $X(t_0)=0.697$, and $Y(t_0)=0.717$, implying $x_0=0.697/\sqrt{\omega_1}$ and $y_0=0.717\sqrt{\omega_1}$.
In Figs.~\ref{fig8}(b) and \ref{fig8}(d) we display the real part of the solution $\psi(kx,t)$ when it acquires the form of a right-going traveling wave $\psi(kx,t)=e^{ikx-i\omega_1(t-t_0)}$ for $t<t_0$. Thus, $f_1(t)=e^{-i\omega_1 (t-t_0)}$, imposing $f_1(t_0)=1$ and $\dot{f_1}(t_0)=-i \omega_1$, while the parameters $(\delta,q)$ and $t_0$ are the same as in Figs.~\ref{fig8}(a) and \ref{fig8}(c). Like the previous case, Fig.~\ref{fig8}(b) shows the time evolution of the traveling wave for $x=0$, while Fig.~\ref{fig8}(d) shows the entire spatiotemporal profile of the traveling wave. It is clearly seen that the transient amplification mechanism also applies for the case of a traveling wave.

\begin{figure}
\begin{center}
\includegraphics[width=1\columnwidth]{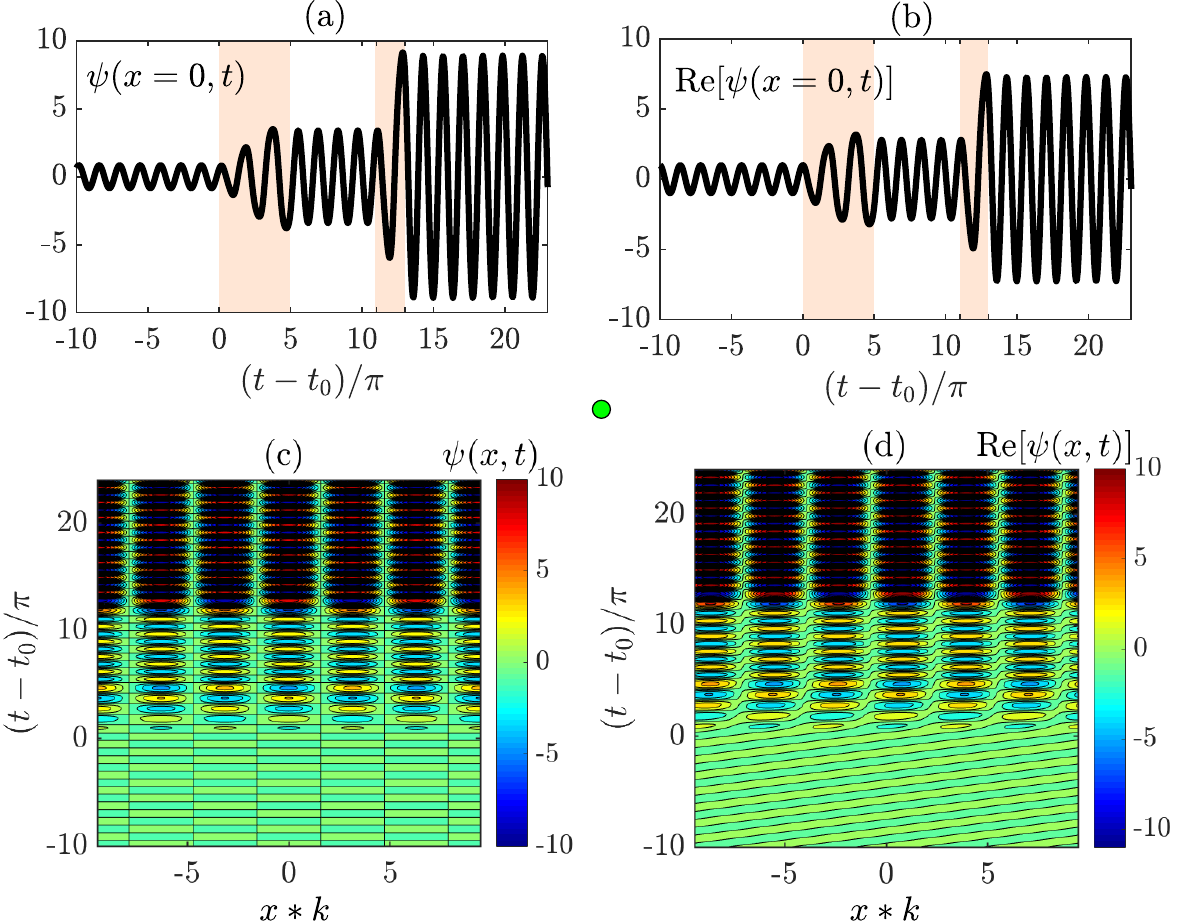}
\caption
{
Similar to Fig.~\ref{fig8}, but using several time interfaces.
The parameters $(\delta,q)$, as well as the initial time $t_0$ and the initial conditions, are the same as in Fig.~\ref{fig8}.
In the orange background, $(t-t_0) \in [0\pi,5\pi]$ and $(t-t_0) \in [11\pi,13\pi]$,  the wave propagates in a medium with time-varying permittivity, while in the white background it is in a constant one.
(a) and (c) correspond to the evolution of an initial standing wave, while 
(b) and (d) correspond to an initial right-going traveling wave.
}
\label{fig9n}
\end{center}
\end{figure}

Next, we consider the wave propagation when a time-varying frequency with more than one time interface is present. Figures~\ref{fig9n}(a)-\ref{fig9n}(d) illustrate the emergence of transient amplification in such a case. Thus, we assume a frequency $\omega^2(t)$ of the form
\begin{equation}
\omega^2(t)
=
\begin{cases}
\delta-2q\cos(2 t_0)=\omega_1^2, & \text{for}~ t \leq t_0,\\
\delta-2q\cos(2 t)=\omega_2^2(t), & \text{for}~ t_0<t \leq t_1, \\
\delta-2q\cos(2 t_1)=\omega_3^3, & \text{for}~ t_1<t \leq t_2, \\
\delta-2q\cos(2 t)=\omega_4^2(t), & \text{for}~ t_2<t \leq t_3, \\
\delta-2q\cos(2 t_2)=\omega_5^2, & \text{for}~ t > t_3. \\
\end{cases}
\end{equation}
The frequency is again continuous at all times, while $(\delta,q)$ are chosen as in the previous example.
The wave field in each time interval is given by
 \begin{equation}
\psi(x,t)
=
\begin{cases}
f_1(t)e^{ikx}, & \text{for}~ t \leq t_0,\\
f_2(t)e^{ikx}, & \text{for}~ t_0<t \leq t_1, \\
f_3(t)e^{ikx}, & \text{for}~ t_1<t \leq t_2, \\
f_4(t)e^{ikx}, &~ \text{for}~ t_2<t \leq t_3, \\
f_5(t)e^{ikx}, &~ \text{for}~ t > t_3. \\
\end{cases}
\end{equation}
 where
 \begin{equation}
 \begin{cases}
f_1(t)=x_0 \cos[\omega_1(t-t_0)] + y_0 \sin[\omega_1(t-t_0)] /\omega_1, \\
f_3(t)=\tilde{x}_0 \cos[\omega_3(t-t_0)] + \tilde{y}_0 \sin[\omega_3(t-t_0)] /\omega_3, \\
f_5(t)= \bar{x}_0 \cos[\omega_5(t-t_0)] + \bar{y}_0 \sin[\omega_5(t-t_0)] /\omega_5, \\
 \end{cases}
\end{equation}
and the functions $f_2(t)$ and $f_4(t)$ satisfy the Mathieu equation with frequencies $\omega_2(t)$ and $\omega_4(t)$, respectively. Furthermore, the initial conditions $x_0$ and $y_0$ and the instants of the time interfaces at $t_0$, $t_1$, $t_2$, and $t_3$ are chosen in the following way: For Figs.~\ref{fig9n}(a) and \ref{fig9n}(c)
we use $x_0=0.697/\sqrt{\omega_1}$, $y_0=0.717\sqrt{\omega_1}$, and 
$t_0=0.32 \pi$, which, as previously discussed, maximize the norm for $f_1(t)$. Then, we set $t_1=5\pi+t_0$, which is the first time that the chosen norm gets its maximum. Subsequently, we set 
$t_2=11\pi+t_0$ since at the end of this time interval, for the given initial conditions $x_0$ and $y_0$, the function $f_4(t)$ is amplified . Finally, we use $t_3=13\pi+t_0$ as an arbitrary choice allowing us to demonstrate the amplification of $f_4(t)$ in a transparent way. 
Note that both $t_{2}$ and $t_3$ are chosen for demonstration reasons and are not derived by an optimization procedure; thus, they are not unique.
This in turn means that the presented amplification in Fig.~\ref{fig9n} is not the optimal one for time $t \geq t_3$. 
Despite this, the illustrated example clarifies that  arbitrary amplification in the stable regime is possible through the appropriate use of several time interfaces.

\subsection{Physical implications}

We note that various wave phenomena found in theory to occur in time-varying platforms have been experimentally validated.
For instance, in Ref. \cite{Fink2016}, the phenomenon of time reflection was observed in a platform with water waves.
By varying the platform periodically in time, the transient amplification of a water wave could be observed as well.

Moreover, phenomena occurring in time-varying media can be observed in electric circuits by introducing appropriate analogies. 
For instance, in Ref. \cite{Engheta2022}, analogies between the electric and magnetic fields and current and voltages were shown.
We note that Ref. \cite{Engheta2022} also numerically illustrated in Fig.~2(b) (even though it was not mentioned) that the electric field of an electromagnetic wave is transiently amplified in a medium with time-varying permittivity [the latter transient effect is due to the non-normality of the propagator matrix, as can be checked using Eq.~(1) and (2)].

Furthermore, it is anticipated that experiments in time-varying optical systems will be done in the near future; see, for instance, the discussion in the introduction of Ref. \cite{Segev2022}.
Therefore, the transient amplification of light could also be observed before long.


\section{Conclusions and discussion}
\label{section9}
In the context of wave propagation in
periodic media, the wave evolution can be described by the Mathieu equation. In this work, we studied the transient amplification features of the stable solutions of the Mathieu equation, which are known to be due to the non-normality of the propagator matrix.
We applied  several methods (the $\epsilon$ pseudospectrum, the Kreiss constant, etc.) classically used in problems of a non-normal nature to quantify this amplification process.
We also took into account the effect of the initial time and showed that the monodromy matrix produces the overall maximum transient gain.
Returning to wave propagation, we showed that this transient amplification of the stable solutions can be used for a controlled increase in wave amplitude as an alternative to the exponential parametric instability. Using different time interface schemes, we showed that arbitrary amplification is possible.

Our work has led to many questions which could be the basis of new studies. 
First of all, the addition of a loss factor is important in view of experimental realization;
then, it is expected that even with asymptotic decay a wave could still experience transient amplification for small times.
Furthermore, it would be of interest to investigate the possible transient amplification experienced by a wave that is scattered through a slab with time-varying permittivity.


\section{Acknowledgements}
The authors would like to thank N. Bakas and P. J. Ioannou for fruitful discussions. I. K. acknowledges financial support from the Institute d'Acoustique - Graduate School of Le Mans and from the Academy of Athens.


\appendix

\section{Norm of the vector ${\xi}$}
\label{AppendixA}

\begin{figure}
\begin{center}
\includegraphics[width=0.9\columnwidth]{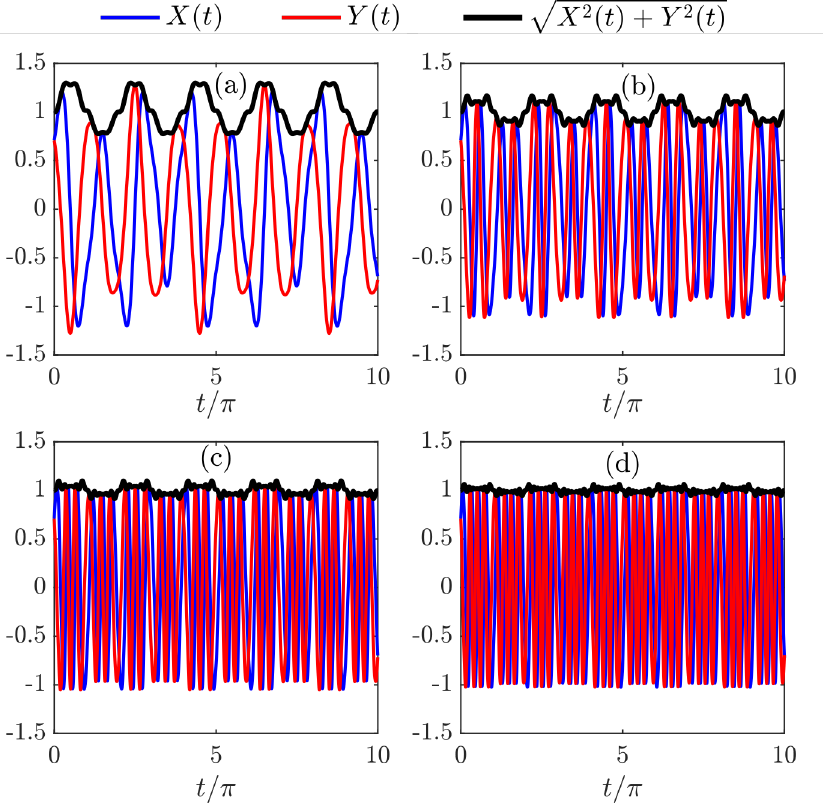}
\caption
{
The variables $X(t)$ and $Y(t)$ and the quantity $\sqrt{|X(t)|^2+|Y(t)|^2}$ as a function of time. 
In all cases $\delta=3q$, with (a) $q=0.82$, (b) $q=4.4$, (c) $q=10.79$, and (d) $q=20.03$. 
All these choices result in an exponent $\gamma$ that is approximately equal to $0.5$.
}
\label{fig9}
\end{center}
\end{figure}

We present in Fig.~\ref{fig9} the norm of the vector ${\boldsymbol{\xi}}(t)$ for four different sets of $(\delta,q)$. In all cases, the parameters lie in the line $\delta=3q$, the exponent $\gamma$  is approximately equal to 0.5, and the initial conditions are $X(0)=Y(0)=1/\sqrt{2}$. 
As the parameters of the Mathieu equation increase, the norm of the vector $\boldsymbol{\xi}$ tends to the constant value $\sqrt{X^2(0)+Y^2(0)}=1$.


\section{Meissner equation}
\label{AppendixB}
\begin{figure}
\begin{center}
\includegraphics[width=1\columnwidth]{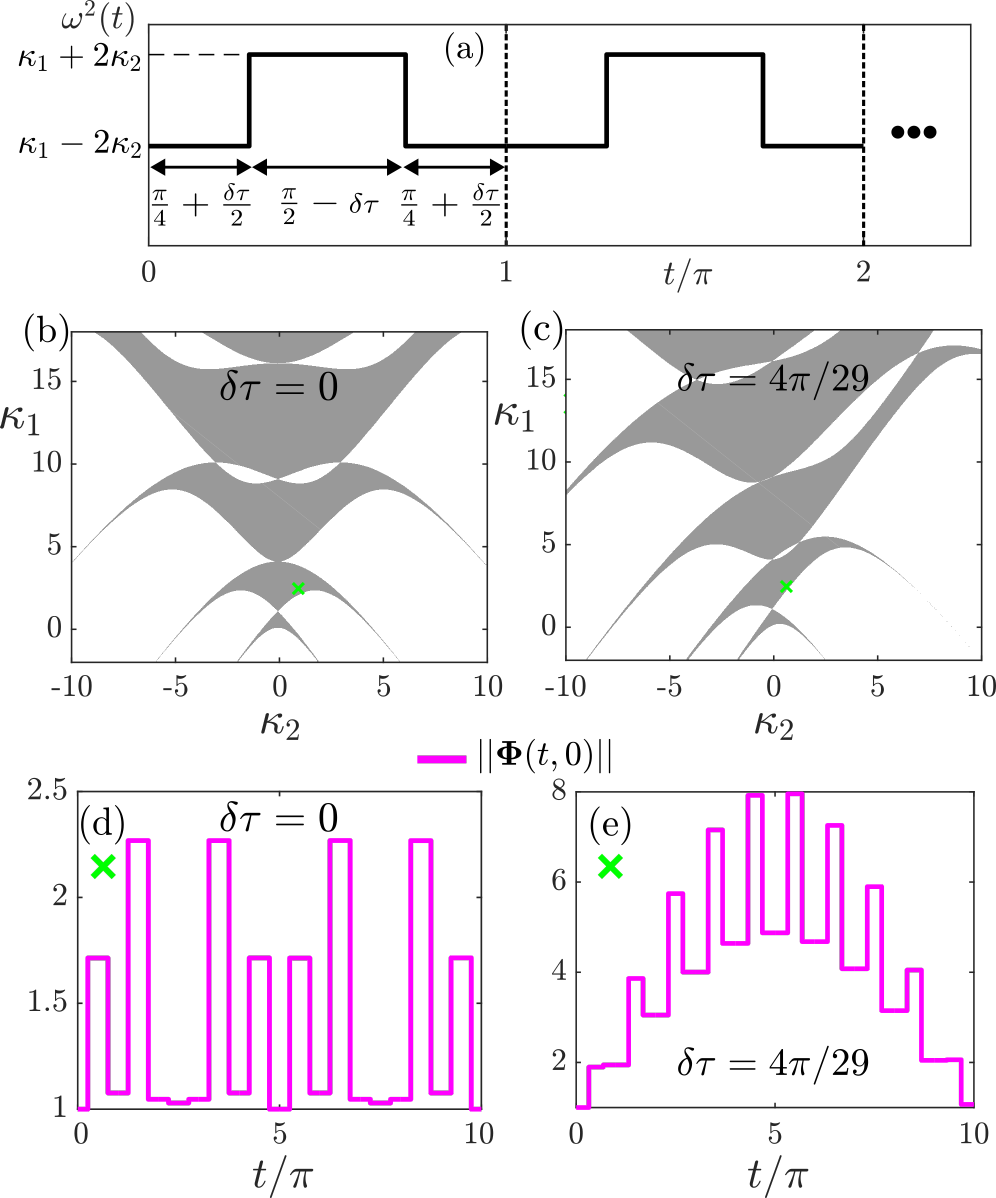}
\caption
{
(a) Piecewise constant frequency $\omega(t)$ of the Meissner oscillator.
Stability diagrams for (b) $\delta \tau =0$ and (c) $\delta \tau =4\pi/29$.
Evolution of the norm of the propagator for (d) $\delta \tau =0$ and (e) $\delta \tau =4\pi/29$.
In both (d) and (e) we set $\kappa_2=0.7585$ and $\kappa_1=3\kappa_2$.
}
\label{fig10}
\end{center}
\end{figure}

We consider here the case of a harmonic oscillator with a piecewise constant time-dependent frequency $\omega(t)$ (the Meissner equation \cite{Koutserimpas2018}), i.e., $\ddot{f}+\omega^2(t)f=0$, varying periodically between the values $\omega_1=\sqrt{\kappa_1-2\kappa_2}$  and $\omega_2=\sqrt{\kappa_1+2\kappa_2}$ with $\kappa_{1,2}$ being constants [see Fig.~\ref{fig10}(a)].
The parameter $\delta \tau$ in Fig.~\ref{fig10}(a) controls the time intervals $\Delta t_{1,2}$ spend in the frequency values $\omega_{1,2}$, where 
$\Delta t_{1,2}=\pi/2\pm \delta \tau$.
When $\delta \tau \neq 0$, the symmetry $\kappa_2 \to -\kappa_2$ is broken, leading to a deformation in the shape of the stability chart.
This is shown in Figs.~\ref{fig10}(b) and \ref{fig10}(c), where two different values of $\delta \tau$ are chosen, $\delta \tau =0$ in Fig.~\ref{fig10}(b) and $\delta \tau =4\pi/29$ in Fig.~\ref{fig10}(c).

In order to quantify the amplification of the stable solutions,
following the same transformation as in Eq.~\eqref{transf},
we find that the monodromy matrix 
\begin{equation}
\Phi(\pi,0)
=
\begin{pmatrix}
\Phi_{11} && \Phi_{12} \\
\Phi_{21} && \Phi_{22}
\end{pmatrix}
\end{equation}
in the transformed variables is given by

\begin{multline}
\Phi_{11}=\cos(\omega_1 \Delta t_1) \cos(\omega_2 \Delta t_2) \\ -\frac{1}{2} \left( \frac{\omega_1}{\omega_2}+\frac{\omega_2}{\omega_1} \right) \sin(\omega_1 \Delta t_1) \sin(\omega_2 \Delta t_2) 
\end{multline}

\begin{multline}
\Phi_{12}=\sin(\omega_1 \Delta t_1) \cos(\omega_2 \Delta t_2) \\ + \left( \frac{\omega_1}{\omega_2}\cos^2(\omega_1\Delta t_1/2)-\frac{\omega_2}{\omega_1} \sin^2(\omega_1\Delta t_1/2) \right) \sin(\omega_2 \Delta t_2) 
\end{multline}

\begin{multline}
\Phi_{21}=-\sin(\omega_1 \Delta t_1) \cos(\omega_2 \Delta t_2) \\ + \left( -\frac{\omega_2}{\omega_1}\cos^2(\omega_1\Delta t_1/2)+\frac{\omega_1}{\omega_2} \sin^2(\omega_1\Delta t_1/2) \right) \sin(\omega_2 \Delta t_2) 
\end{multline}

\begin{equation}
\Phi_{22}=\Phi_{11}.
\end{equation}

In Figs.~\ref{fig10}(d) and \ref{fig10}(e) we present the norm of the propagator matrix $\mathbf{\Phi}(t,0)$ for the same point in the stability chart (green cross) but for the two different $\delta \tau$ used in Figs.~\ref{fig10}(b) and \ref{fig10}(c).
In both cases we are in the stable region, but transient amplification is observed.

\begin{thebibliography}{99}
\bibitem{commentnonlinear} Note that the linearization of a nonlinear dynamical system around a periodic solution leads in many cases to Hill’s differential equation (for instance, the linearization of the Duffing equation leads to the Mathieu equation), suggesting a connection between nonlinear evolution and transient amplification.
\bibitem{comment1} Note that the variables $X$ and $Y$ are real only when the parameter $\omega^2(t)$ is positive, that is, inside the cone $\delta=2|q|$ is shown in Fig.~\ref{fig1}(a). At the boundary of this cone as well as in the outside region, $\omega(t)$ becomes zero within each period $\pi$, suggesting that the variable  $Y$ acquires singularities. As a result, another suitable transformation should be considered that describes the amplification features correctly in this region. However, there is no reason for the amplification features to be qualitatively different in these two regions since the initial variables $f$ and $\dot{f}$ which express the physical properties of the system change smoothly across the whole stability chart. Therefore, we restrict our analysis to only the region located inside the cone $\delta=2|q|$, where the norm of the vector ${\boldsymbol{\xi}}(t)$ properly quantifies  the amplification features.
\bibitem{comment2} Here we are considering a case where the propagator is periodic and its maximum is taken over one period.
We also tried cases of nonperiodic evolution over a very long time, and we were unable to see behavior different  from the one described.
\bibitem{comment3}  We make the following remarks regarding the intervals that the two times $t$ and $t_0$ must scan in order to get the desired maximum numerically: First, even though the quantity $\omega(t)$ is periodic with period $\pi$, it is sufficient for the time $t_0$ to scan the interval $[0,\pi/2]$ (see Fig.~\ref{fig6}). Second, when the exponent $\gamma$ is a rational number it is sufficient for $t$ to scan the interval $[0,\pi m_2]$ (recall Fig.~\ref{fig3}) in order to reach the maximum amplification.
\end{thebibliography}
\end{document}